\documentclass[journal=jpccck,manuscript=article]{achemso}
\usepackage[T1]{fontenc} % Use modern font encodings
\usepackage{amsmath}
\usepackage{amssymb}
\usepackage{graphicx}% Include figure files
\usepackage{dcolumn}% Align table columns on decimal point
\usepackage{bm}% bold math
\usepackage[position=top]{subfig} % caption=false,
\usepackage{nameref}
\usepackage{xcolor, soul}

\graphicspath{ {figs/} } % location of images

\author{Hector Lopez-Rios}
\affiliation{Department of Materials Science and Engineering, Northwestern University}
\altaffiliation{Equal contribution}
\author{Ali Ehlen}
\affiliation{Applied Physics Program, Northwestern University}
\altaffiliation{Equal contribution}
\author{Monica Olvera de la Cruz}
\affiliation{Department of Materials Science and Engineering, Northwestern University}
\alsoaffiliation{Applied Physics Program, Northwestern University}
\email{m-olvera@northwestern.edu}
\phone{+1 (847) 491-7801}

\title{Delocalization Transition in Colloidal Crystals}

\abbreviations{}

\begin{document}

%\begin{tocentry}
%
%Some journals require a graphical entry for the Table of Contents.
%This should be laid out ``print ready'' so that the sizing of the
%Inside the \texttt{tocentry} environment, the font used is Helvetica
%8\,pt, as required by \latin{Journal of the American Chemical
%The surrounding frame is 9\,cm by 3.5\,cm, which is the maximum
%permitted for  \latin{Journal of the American Chemical Society}
%graphical table of content entries. The box will not resize if the
%content is too big: instead it will overflow the edge of the box.
%
%This box and the associated title will %always be printed on a
%separate page at the end of the document.
%
%\end{tocentry}

\begin{abstract}
  Sublattice melting is the loss of order of one lattice component in binary or ternary ionic crystals upon increase in temperature. A related transition has been predicted in colloidal crystals. To understand the nature of this transition, we study delocalization in self-assembled, size asymmetric binary colloidal crystals using a generalized molecular dynamics model. Focusing on BCC lattices, we observe a smooth change from localized-to-delocalized interstitial particles for a variety of interaction strengths. Thermodynamic arguments, mainly the absence of a discontinuity in the heat capacity, suggest that the passage from localization-to-delocalization is continuous and not a phase transition. This change is enhanced by lattice vibrations, and the temperature of the onset of delocalization can be tuned by the strength of the interaction between the colloid species. Therefore, the localized and delocalized regimes of the sublattice are dominated by enthalpic and entropic driving forces, respectively. This work sets the stage for future studies of sublattice melting in colloidal systems with different stoichiometries and lattice types, and it provides insights into superionic materials, which have potential for application in energy storage technologies.     
\end{abstract}

\section{Introduction}

Binary colloidal systems, which have interspecies attraction and intraspecies repulsion, have been shown to self-assemble into a wide variety of binary lattices.\cite{Shevchenko2006,Hynninen2006,Bodnarchuk2010,Hueckel2020,Oh2020} Generally, if the two colloid species are of sufficiently different sizes, the larger colloids will form a lattice while the smaller colloids occupy interstitial sites.\cite{Leunissen2005, Filion2011, VanDerMeer2017, Girard2019} In these size asymmetric colloidal systems, many cubic and non-cubic crystals have been detected, including a Frank-Kasper phase.\cite{Girard2019} However, under certain conditions, the small particles may delocalize and roam around the crystal while the large particles remain in lattice sites; this is called sublattice melting. Previously, this behavior had been seen primarily in atomic systems, in materials termed superionics\cite{Hull2004, Schommers1977, Tatsumisago1991}, where one ionic species delocalizes while the other stays fixed in a lattice. However, recent work has demonstrated sublattice melting in assemblies of hard spheres under pressure,\cite{Filion2011, VanDerMeer2017} oppositely charged colloids with a Debye-H\"{u}ckle potential,\cite{Lin2020} and colloids functionalized with sticky DNA chains.\cite{Girard2019} The surprising loss of order of only the sublattice also resembles behavior found in metals. In this analogy, the small particles map to delocalized electrons and the large particles to fixed nuclei. Given the unique physical nature of this phenomenon in colloidal systems and the seeming generality of the colloidal crystals that exhibit it, we seek to understand the origin of colloidal sublattice melting using a simplified molecular dynamics (MD) model, which can provide insight into a range of systems.

To calculate reliable thermodynamic and physical quantities of delocalized systems, we developed a scalable MD model. This simplified model enables us to generalize previous work that predicted delocalization in systems of DNA-functionalized gold nanoparticles,\cite{Girard2019} where the interactions between colloid species were due to DNA hybridization, which is directional and specific. However, the experimental design also included additional free DNA chains that may have acted as depletants. To avoid complications related to DNA hybridization and to explore the generality of the phenomenon, the pairwise interactions in our model are isotropic and short-range.

The generality of this model also enables us to apply it to a wide variety of systems. This encompasses, for example, nanodots with thiols and end terminal attractive groups,\cite{Donakowski2010, Harris2016} functionalized nanoparticles with light activated interactions,\cite{Kalsin2006,Klajn2007} and nanocomposite tectons.\cite{Zhang2016,Santos2019a,Santos2019b} In fact, nanocomposite tectons would be an ideal system for experimental verification of this study, because the parameters of the system reported in the present work can correspond to metallic nanoparticles functionalized with hydrocarbon chains with short ranged and strong complementary molecular binding pairs. Lastly, with this model, we can start to address questions that have been posed about sublattice melting in superionic materials \cite{Wang2015,Muy2018} such as the origin of the sublattice melting transition. However, superionic materials are constrained by the requirement of charge neutrality per unit cell, but colloidal crystals (and this model) have no such constraint.

In this paper, we study the localized-to-delocalized transition in functionalized, size asymmetric colloidal crystals. We explore the order of this transition with respect to temperature and by varying the number of chains per small particle (4, 6, 8, and 10 chains per small particle). We focus on a system composition of 6 small particles per large particle ("6:1 ratio"), because at this composition, the large particles form a stable body centered cubic (BCC) lattice over a wide temperature range. Though other compositions exhibit interesting symmetry changes with temperature and number of chains, we use the 6:1 ratio to study the nature of the localized-to-delocalized transition without the added complexity of a change in the large particle lattice. 

\begin{figure}
    \centering
    \includegraphics{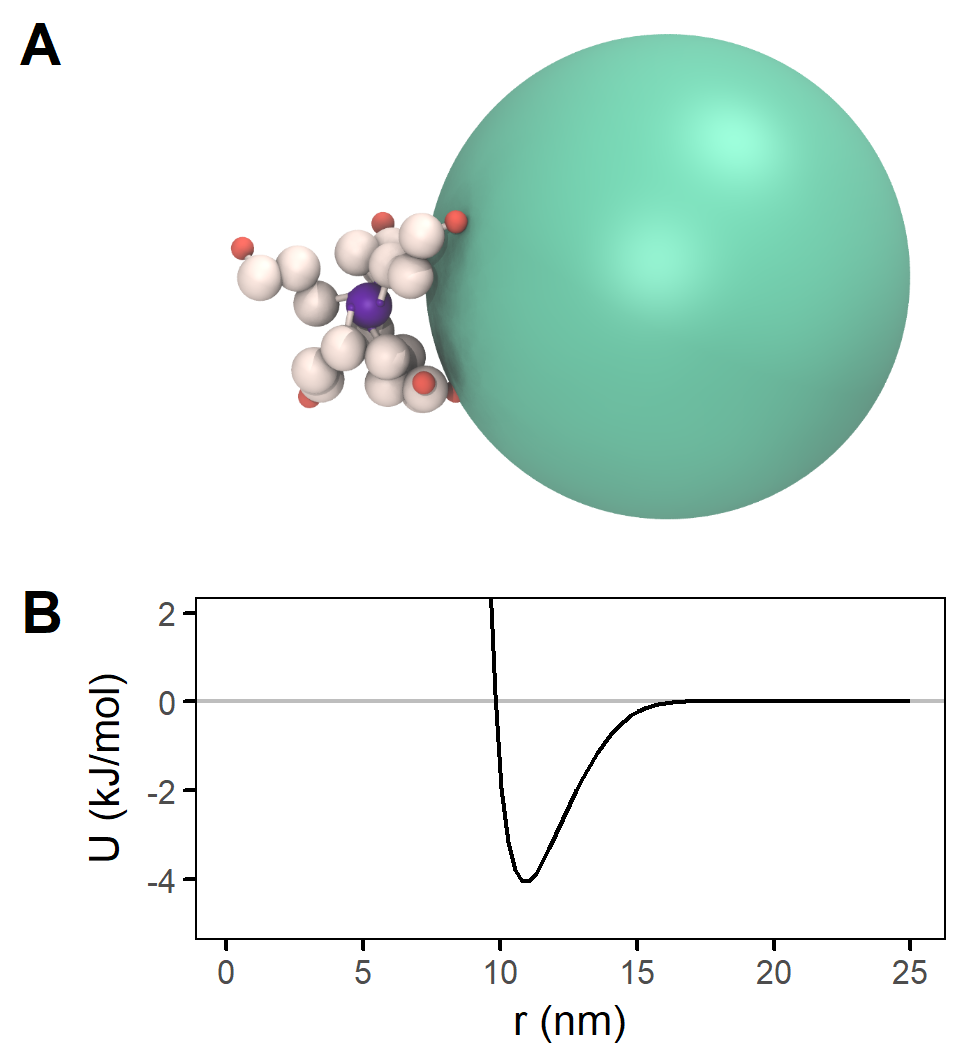}
    \caption{\label{fig:model} The simplified model. 
    \textbf{(A)} The smaller colloid (purple) functionalized with chains (white) and larger colloid (turquoise) in our system, to scale. All beads have excluded volume based on their radii, and there is an attractive interaction between the large particles and the interactive ends of the small particle chains (orange). 
    \textbf{(B)} Pair potential $U(r)$ between the centers of the large particles and the interactive chain ends. The high-energy region $r < 10$ nm represents excluded volume interactions, and the small potential well accounts for the attractive interaction. The value of $U$ at the minimum is -4.1 kJ/mol.}
\end{figure}

An image of the model is shown in Figure \ref{fig:model}A. The larger colloidal species is represented by a single sphere (shown in turquoise). The smaller species is represented by a small sphere (shown in purple) grafted with a variable number of self-avoiding chains (shown in white). The only interactions in the system are excluded volume between all beads, using a Weeks-Chandler-Andersen (WCA) potential, and a generalized, attractive potential between the large particles and the interactive ends of the chains (shown in orange, referred to here as "interactive ends"); see Figure \ref{fig:model}B. There is asymmetry in the interaction potential, as the range of the attractive potential is half of the diameter of the large particles. We chose to further simplify the system by representing the large species as spheres without explicit functionalized chains. This choice is consistent with colloidal systems that have previously shown sublattice melting, because these systems' large particles were either spherical\cite{Filion2011, Lin2020} or densely enough grafted with polymer chains\cite{Girard2019} that a spherical potential is a reasonable approximation. However, the small particles cannot be modeled as spheres due to their higher curvature and therefore lower packing density of grafted chains. When chains are omitted and the interaction potential between small and large particles is modelled with spherical potentials, mostly FCC crystals are obtained.\cite{Lin2020,Filion2011} This may be because explicitly representing grafted chains on the small particles also enables spatially anisotropic interactions between the small and large particles. These can occur when chains bundle together in configurations analogous to hybridization electron orbitals present in covalent bonding called skyrmions.\cite{Travesset2017, Santos2019a} The concept of skyrmions has proved useful in explaining the appearance of non-close packed functionalized colloidal crystal structures.

With this model, we find that both size and interaction range asymmetries are needed for delocalization to occur. The passage from localized-to-delocalized small particles is fully continuous, closely related to diffusion of the small particles, and enhanced by the vibrational entropy of the large particle lattice. This continuous behavior arises from a competition between enthalpic and entropic driving forces. Here enthalpic contributions can be understood through analysis of the interaction energy landscape between the large particles and interactive beads of the small particles. Entropic contributions arise from the vibrations of the large particle lattice. 

The rest of the paper is organized as follows. We begin by describing the MD simulations, as well as a theoretical model used for the free energy analysis of these crystals. We continue with a symmetry and energy analysis of relevant aspects of the BCC sublattice and its energy landscape. We then describe an analysis of the temperature-dependent thermodynamic and physical properties such as lattice parameter and specific heat per particle. We conclude by analyzing the importance of lattice vibrations as the driving force for both delocalization and lattice expansion for these crystals.

\section{Methods}
\label{sect:methods}

\subsection{General description of the MD model}

There are two types of pair interactions between the beads in the system. First, all beads have excluded volume interactions with each other through the WCA potential. Second, there is an attractive interaction between the interactive end of the chain and the large particles. That interaction is in the form of a Gaussian potential and is shown graphically in Figure \ref{fig:model}B and mathematically here:
\begin{align}
\label{eq:pair-potential}
    U_\text{pair} (r)
        =& 
        \begin{cases}
            U_\text{WCA} (r) + U_\text{Gaussian} (r) & r \le r_\text{cutoff}   \\ 
            0 & \text{otherwise}
        \end{cases}
\end{align}
where
\begin{align}
    U_\text{WCA} (r)
        =& ~
        4 
        \left( 
            \left( \frac{\sigma}{r} \right)^{12}
            -
            \left( \frac{\sigma}{r} \right)^{6}
        \right)
        -
        4 
        \left( 
            \left( \frac{\sigma}{2^{1/6} \sigma} \right)^{12}
            -
            \left( \frac{\sigma}{2^{1/6} \sigma} \right)^{6}
        \right)
        &&\text{ for } r \le 2^{1/6} \sigma
        \\
        \label{eq:gauss}
    U_\text{Gauss} (r)   
        =&
        -\varepsilon
        e^{-\frac{1}{2} \left( \frac{r}{\sigma_\text{gauss}} \right)^2}
        &&\text{ for } r \le r_\text{cutoff} 
\end{align}
where $r$ is the distance between the centers of the large particle and the interactive end bead of the small particle chains; $\sigma = \sigma_\text{large} + \sigma_\text{int. bead}$, the sum of the radii of the large particle and the interactive bead; $\varepsilon$ is a (positive valued) parameter that determines the strength of an individual large particle-interactive bead interaction; and $\sigma_\text{gauss}$ is a parameter that determines the range of $U_\text{Gauss} (r)$. As usual, the WCA potential is cut off at $2^{1/6} \sigma$ and shifted such that $U_\text{WCA}$ is zero at the cutoff, that is $U_\text{WCA}(r=2^{1/6} \sigma) = 0$. The value for $r_\text{cutoff}$ was selected such that $U_\text{Gauss} (r)$ has safely decayed to near zero by $r=r_\text{cutoff}$. We also used the HOOMD-blue \texttt{xplor} option which adds a subtle smoothing near $r_\text{cutoff}$ such that the $U_\text{Gaussian}$ decays smoothly to zero.\footnote{See \texttt{md.pair.pair} documentation:\\ \texttt{https://hoomd-blue.readthedocs.io/en/stable/module-md-pair.html}}

These parameters can be adjusted such that the system resembles interactions between two colloid species of choice. Additional parameters may vary are: particle size, number of chains on each small particle, temperature, system composition (ratio of small:large colloids in the simulation box), and length and stiffness of the chains on the small particles. The properties chosen for the study in this paper are listed in the next section.

\subsection{Parameters and simulation scheme}\label{sect:params}

\begin{table}
    \centering
    \subfloat[\label{subtab:params-fix}]{
        \begin{tabular}{|cc|}
            \hline 
            \textbf{Parameter} & \textbf{Value} \\
            \hline 
            $\sigma_\text{large particle}$ & 10.5 nm \\
            $\sigma_\text{small particle center}$ & 1.0 nm \\
            $\sigma_\text{chain bead}$ & 1.0 nm \\
            $\sigma_\text{interactive chain end bead}$ & 0.5 nm \\
            $\varepsilon$ & 70 kJ/mol \\
            $\sigma_{gauss}$ & 4.8 nm \\
            $r_\text{cutoff}$ & 8.4 nm \\
            \# non-interactive beads/chain & 3 \\
            % large particle mass & 3 R  \\
            % small particle mass & 3 R \\
            % chain bead mass & 1 R \\
            \hline
        \end{tabular}
    }
    ~~ 
    \subfloat[\label{subtab:params-var}]{
        \begin{tabular}{|cc|}
            \hline 
            \textbf{Parameter} & \textbf{Value} \\
            \hline 
            small:large particle ratio & 6:1 \\
            temperature & $k_B T = 0.8 - 2$ kJ/mol \\
            \# chains/small particle & 4, 6, 8, 10 \\
            \hline
        \end{tabular}
    }
    \caption{ \label{tab:params}
    Parameters used in the present study.     
    \textbf{(a)} Fixed parameters ($\sigma$ is radius). With these, the system resembles a binary system of weakly interacting chain-grafted colloids.
    \textbf{(b)} Variable parameters. Changing these allows us to explore properties of the system.
    }
\end{table}

We chose parameters for the interaction of our particles to generalize the short ranged attractive potential found in self-assembled DNA functionalized colloidal crystals\cite{Knorowski2011, Li2012, Biancaniello2005, Rogers2011}. DNA functionalized colloids interact by forming hydrogen bonds between the single stranded DNA at the ends of the grafted chains. Using the parameters in Table \ref{tab:params}A, at $T^*=1$, the potential well shown in Figure \ref{fig:model}B has a depth of -4.1 $k_B T$, which is approximately the binding energy of hydrogen bonding in single-stranded DNA (3 - 6 $k_{B} T$ \cite{Biancaniello2005,Rogers2011}). However, given the general nature of our model, other forms of interactions found in functionalized colloidal crystals, such as dispersion interactions, can be represented with this model. Additionally, we fixed a particle size asymmetry that is in the regime in which binary solids form interstitial solid solutions (ISSs), where the smaller species occupies interstitial sites of the large species lattice. For example, in metallic binary alloys, one of the Hume-Rothery rules\cite{Hume-Rothery1966} require atomic size asymmetries where the smaller species size is $\leq 0.4$ the size of the larger species in order to form ISSs. For functionalized binary colloidal particles, it was experimentally demonstrated\cite{Girard2019} that ISSs were formed only for particle diameter ratios of 10 to 1.4 nm, while they were not formed when the smaller particles where larger.

For this study, we ran simulations of colloidal systems at different temperatures and number of grafted chains per small particle, as detailed in Table \ref{tab:params}B. Varying both temperature and number of chains allows us to explore a wide range of system states. Changing the number of chains per small particle changes the total attraction strength between small and large particles, as well as the symmetry of available chain configurations. Additionally, because the attractive interaction is simple (Equation \eqref{eq:gauss}), the system's behavior is determined by the ratio $\varepsilon/k_BT$. Therefore, by varying temperature, we are also effectively examining the range of behavior that would appear if we instead varied interaction strength. 

All simulations were run using HOOMD-blue version 2.5.1\cite{Anderson2008, Glaser2015} in the NPT ensemble with periodic boundary conditions at near-zero pressure (207 Pa, which is $\sim$ 2\% of atmospheric pressure). Using a pressure very close to zero enables us to attribute the observed crystal assembly to the interactions between colloids, rather than an external pressure.\cite{Girard2019} Additionally, during the NPT portion of the run, the box was allowed to fluctuate in size and shape, which enabled lattices that were initialized in one crystal structure to relax into another if it was favorable to do so. 

The full simulation scheme is as follows: we started the simulations in various initial lattice configurations (BCC, SC, FCC, BCT) with 6x6x6 unit cells in the simulation box. The simulations were then equilibrated, thermalized, and depressurized to their final pressure. This initial sequence lasted 312 ns. Then, the simulations were run in the NPT ensemble for an additional 8.44 $\mu$s. For analysis, the first 1.38 $\mu$s were considered to be an equilibration period and not included in calculation of properties. Therefore, analysis of the simulations was conducted on the last 7.37 $\mu$s.

System topology for the simulation was built using Hoobas, \cite{Girard2019a} analysis was done in Python using MDAnalysis \cite{Michaud-Agrawal2011, Gowers2016} and R, visualization of the simulation was done in VMD \cite{Humphrey1996} with the GSD plugin\footnote{See HOOMD-blue GSD plugin for VMD at
\texttt{https://github.com/mphoward/gsd-vmd}} using the internal Tachyon ray-tracing library \cite{Stone1998} (see Figure \ref{fig:model}A), and scientific plotting and calculation of isosurfaces and 3-dimensional densities (see Figures \ref{fig:loc-vs-deloc_deloc} and  \ref{fig:fixedlatt_loc-vs-deloc-images}) was done in Mayavi.\cite{Ramachandran2011}

\subsection{Theoretical free energy of the exact soluble model}

The theoretical model described in Section \textit{\nameref{sect:theory}} is derived by calculating the energetic environment of one interactive bead in one unit cell of a fixed BCC lattice of large particles. That is: 
\begin{align}
\label{eq:partitionfunction}
    \nonumber
    Z(a, T) &= \int \int e^{- U_\text{end}(\vec{r},\vec{p};a)/k_BT}~d\vec{r}~ d\vec{p} \\
    Z(a, T) &= \left( 2 \pi m k_BT \right)^{3/2} \int e^{- U_\text{potential}(\vec{r};a)/k_BT}~d\vec{r} 
\end{align}
where $U_\text{end}(\vec{r},\vec{p};a)$ is the energy associated with the particles in one unit cell with lattice parameter $a$ and an interactive end with position $\vec{r}$ and momentum $\vec{p}$. The position of the interactive bead $\vec{r}$ is integrated over one unit cell and its momentum $d\vec{p}$ is integrated over all real numbers (this Gaussian integral is known from the ideal gas partition function). The integral has been simplified using the definition of energy $U_\text{end}$ as:
\begin{align*}
    U_\text{end}(\vec{r},\vec{p};a)
    &=
    \frac{\vec{p}^2}{2m}
    +
    U_\text{potential}(\vec{r};a) \\
U_\text{potential}(\vec{r};a)
    &=
    \sum_n U_\text{pair}(|\vec{r} - \vec{R}_n|;a)
    +
    \sum_{j<k} U_\text{WCA} (|\vec{R}_j - \vec{R}_k|)
\end{align*}
where $U_\text{pair}(\vec{r};a)$ is the pair potential between a large particle and an interactive bead, as defined in Equation \ref{eq:pair-potential}, and the sum is taken over all large particles that could influence the energy of an interactive bead at $\vec{r}$ ($\vec{R}_n$ indicates the position of the $n$th large particle). In this case, we include 15 large particles: all 9 pictured in the BCC cell in Figure \ref{fig:BCC_sites}A, plus the large particles in the center of all 6 non-diagonal adjacent unit cells. The range over which $U_\text{pair}(r;a)$ is nonzero in this model is short enough such that this captures all interactions. $U_\text{WCA}(r)$ is the WCA potential between large particles; this term becomes important when $a$ approaches the diameter of the large particles.

We then numerically integrate Equation \ref{eq:partitionfunction} to find the partition function, and we can set up equations to calculate any statistical mechanical quantity that can be found with that result. For example, to calculate the average interaction energy between small and large particles, we numerically evaluate the following (assuming $a$ is large enough that $U_\text{WCA}(r)$ can be neglected):
\begin{align}
    \label{eq:theory-int-energy}
    \left< U_\text{potential}(a, T) \right>
    &= 
    \frac{1}{Z(a, T)} \left( 2 \pi m k_BT \right)^{3/2} 
    \int
        \left(
            \sum_n U_\text{pair}(|\vec{r} - \vec{R}_n|;a)
        \right)
        e^{-U_\text{potential}(\vec{r};a)/k_BT}~d\vec{r} 
\end{align}
The partition function is also used to calculate free energy using: \begin{equation*}
    F(a, T) = - k_BT \ln \left( Z(a, T) \right)
\end{equation*}

This model enables us to understand how the BCC energy landscape impacts system behavior, despite its simplicity. For example, it does not include lattice fluctuations. However, the lack of lattice fluctuations impacts the variance but not the mean of predicted energy values (we have seen this trend when comparing the mean and variance of the interaction energy between the fixed and fluctuating lattice cases). 

Additionally, this model does not include particles other than the lattice and a single interactive bead. This is a sufficient approximation because the interaction between the small and large particles is more significant than the interaction between small particles. That is particularly true when small particles have fewer chains, because the small particles interact with 4 large particles when they sit at BCC tetrahedral sites. When there are 4-6 chains on each small particle, each chain is, on average, attracted to one of the 4 nearby but physically separated potential wells (see Figures \ref{fig:BCC_sites}B and \ref{fig:BCC_sites}C). Therefore, their excluded volume interactions don't substantially impact their average energy values, and agreement between theory and simulation is stronger for systems with fewer chains per small particle. However, as described later, the theory's lack of bond constraints does matter. In simulation, the bonds in small particle chains don't allow interactive beads to access the lowest-energy part of the unit cell's potential wells. However, this appears to simply scale the average energy of the interactive beads, especially, as noted, for systems with fewer chains. 

Lastly, note that the lattice parameter and temperature are both inputs to this partition function. It is possible that this formulation could predict some lattice expansion as a function of temperature. However, because of the differences in average location of the interactive bead between theory and simulation (due to bond constraints), we do not believe that this will be a quantitative prediction for properties of a fluctuating lattice simulation. Despite this, this theory can provide a sense of how much the lattice vibrations contribute to certain properties of a system where they are present.

\section{Results and Discussion}

\subsection{6:1 systems form BCC lattices with small particles localized at tetrahedral sites}
\label{sect:6:1BCC}

\begin{figure}
    \centering
    \includegraphics{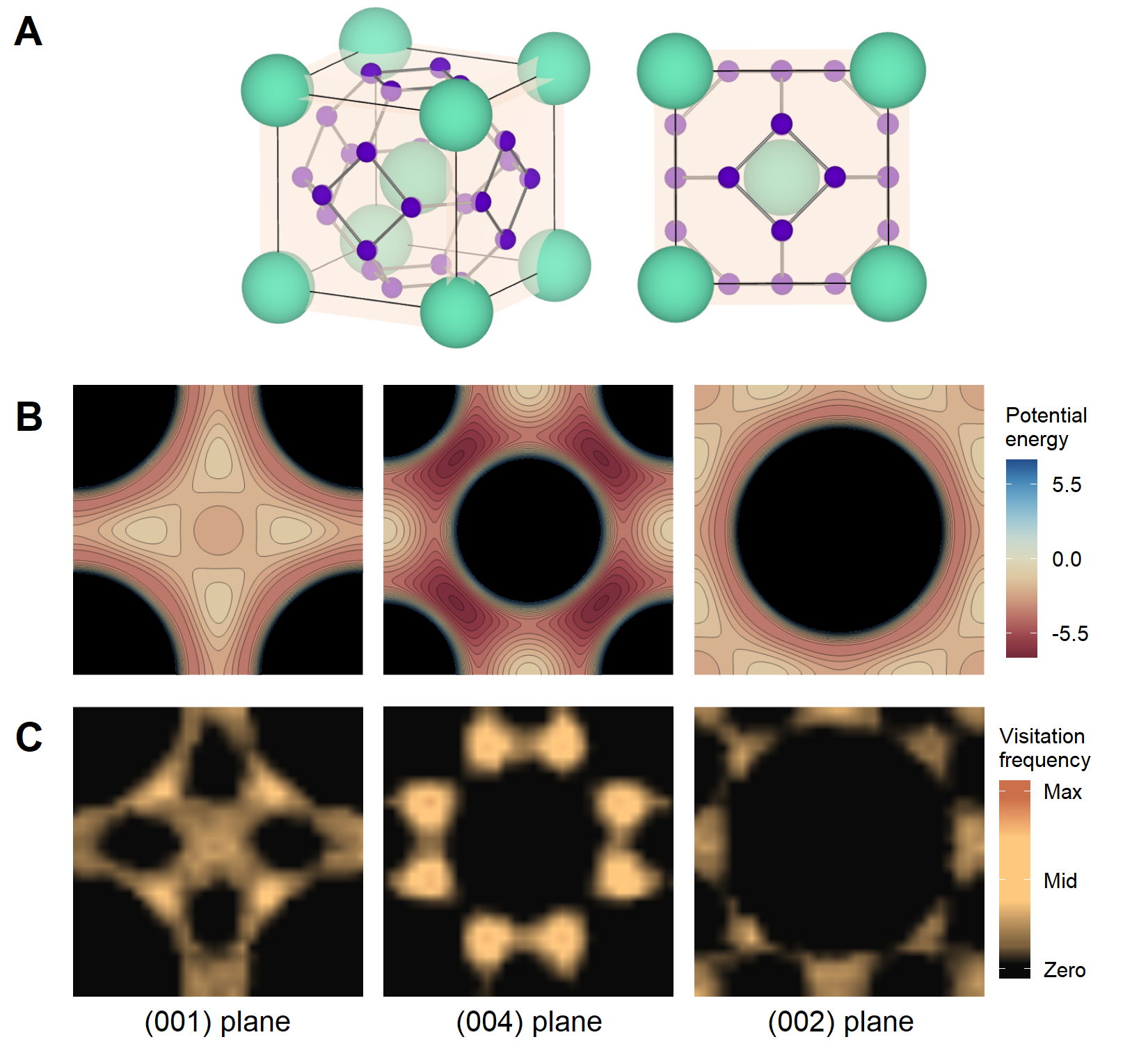}
    \caption{\label{fig:BCC_sites} 
    \textbf{(A)} BCC lattice sites (turquoise) and tetrahedral interstitial sites (purple) of an ideal BCC lattice. Connections between nearest-neighbor tetrahedral sites are shown as visual guides. 
    \textbf{(B)} The potential energy landscape in different planes of one interactive end, based on its interaction potential with the large particles, in one BCC unit cell. Deeper red indicates negative values (more favorable energetic interaction), yellow indicates values around zero, and dark blue indicates positive values (unfavorable interactions; the location of large particles is shown in black). 
    \textbf{(C)} The probability distribution of the interactive beads on different planes for the case of 6 chains at $T^*=0.9$. Comparing this to (B), interactive bead probability is highest in areas with the most favorable energetic interactions.
    }
\end{figure}

For each value of chains per small particle, 6:1 systems form BCC lattices over a wide temperature range. This is consistent with findings of Girdard, et al.\cite{Girard2019} with respect to their 6:1 systems. At temperatures below this range, we observe formation of other crystal lattice types, and at higher temperatures, we observe liquid or gas phases; see SI for more information on determining BCC stability. At lower temperatures within the BCC range, the large particles sit at BCC lattice points and the small particles localize at the BCC tetrahedral sites, also known as 12d Wyckoff positions; these are shown in Figure \ref{fig:BCC_sites}A. The location of the tetrahedral sites means that each small particle can interact with four large particles simultaneously.

An analysis of the symmetry and energy associated with the tetrahedral sites reveals why small particles localize there. The potential energy of interaction between large particles and the interactive bead at the end of each chain can be seen in Figure \ref{fig:BCC_sites}B. Dark red indicates negative interaction energy and defines the areas most favorable for the interactive ends to occupy. Conversely, the lighter areas indicate an interaction energy of approximately zero. There are four nearly zero energy sites per face, visible in the (001) plane image in Figure \ref{fig:BCC_sites}B. These are the tetrahedral sites. This suggests that the small particle centers localize at the tetrahedral sites because this enables the interactive ends to access the most energetically favorable regions of the unit cell. Tetrahedral structures have also been observed experimentally. The formation of distorted tetrahedral structures between size asymmetric colloids has been reported within a specific size asymmetry range (which does not include the dimensions of our system)\cite{Schade2013}. The experimental tetrahedral clusters, mediated by short ranged but strong potentials (both electrostatic and DNA hybridization), were explained using entropic principles. Here, enthalpy seems to be the predominant driving force for the formation of these BCC crystals.       

The energy landscape show in Figure \ref{fig:BCC_sites}B is a good predictor of the locations of particles in simulation. Figure \ref{fig:loc-vs-deloc_deloc}A shows the probability density of the small particle centers in a single BCC unit cell at low temperature. The small particles are clearly localized at the tetrahedral sites. Additionally, Figure \ref{fig:BCC_sites}C shows the probability density of the interactive ends in a low temperature simulation. The location of the highest density regions aligns well with the lowest energy positions in Figure \ref{fig:BCC_sites}B. A notable exception is that the limited reach of the chains in simulation does not allow the interactive beads to reach the bottom of each potential well. 

Lastly, the 6:1 number ratio between small and large particles allows the tetrahedral sites to be exactly filled. This is because there are 2 lattice points (large particles) and 12 tetrahedral sites (small particles) per BCC unit cell. A lower ratio would produce vacancies in tetrahedral sites; in those cases, we observe hopping of small particles between sites. A larger ratio results in more small particles than available tetrahedral sites; in those cases, interstitial defects are prominent and full localization is not possible. Studying the 6:1 system allows us to focus on the properties of the localized-to-delocalized transition by avoid confounding factors introduced by vacancy hopping or symmetry change.    

\subsection{The localized-to-delocalized transition is smooth and its onset depends on interaction strength}
\label{sect:transition}

\begin{figure}
    \centering
    \includegraphics{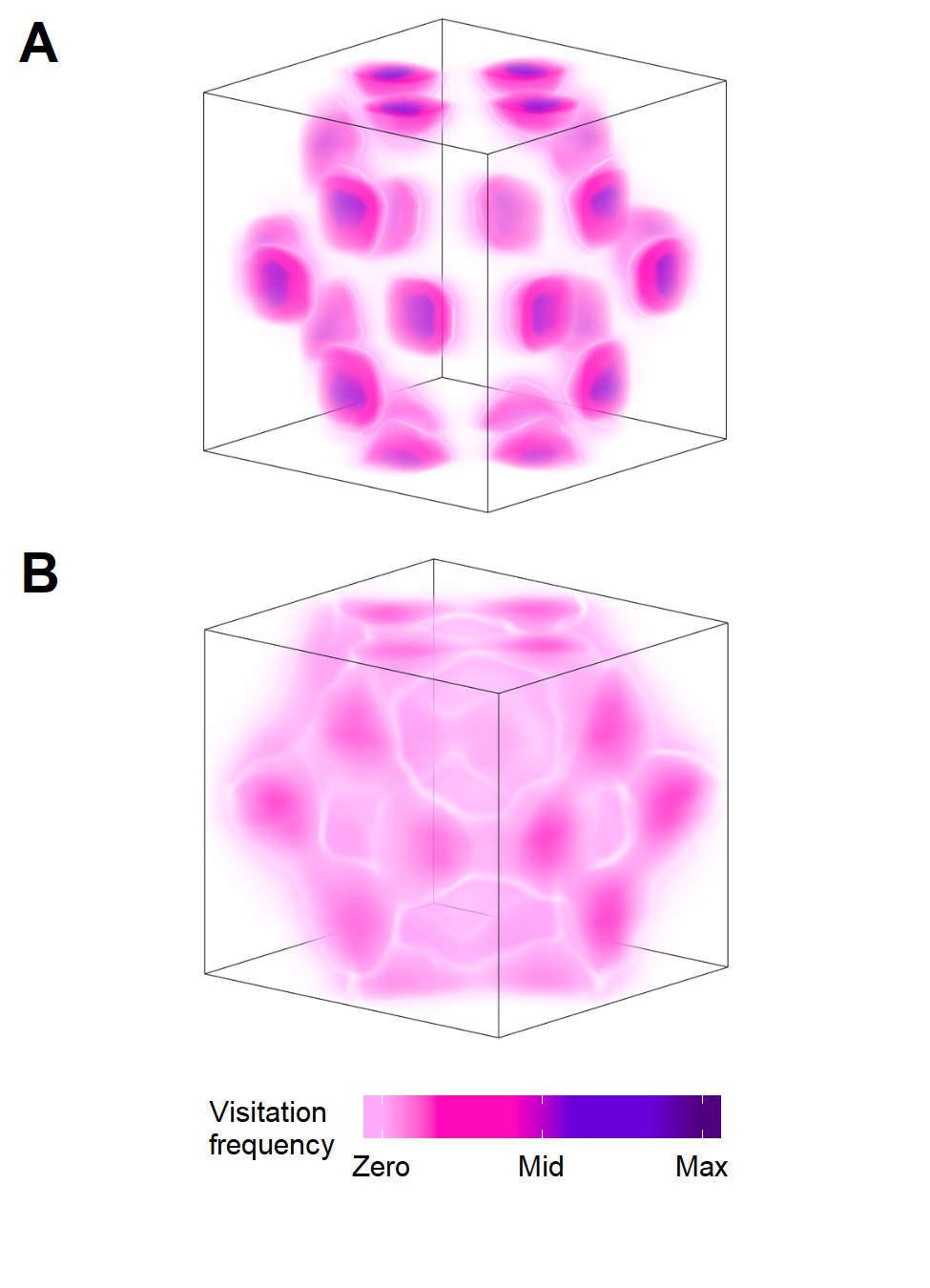}
    \caption{\label{fig:loc-vs-deloc_deloc} Visitation frequency of (centers of) small particles in one unit cell of localized and delocalized systems. Small particles have 6 chains, and the maximum of the visitation frequency is 0.0115.
    \textbf{(A)}  $T^* = 0.9$. Small particles are localized on the tetrahedral sites of the BCC lattice. 
    \textbf{(B)} $T^* = 1.6$. Small particles are delocalized. They favor the tetrahedral sites of the BCC lattice but also roam around the crystal. 
    }
\end{figure}

\begin{figure}
    \centering
    \includegraphics{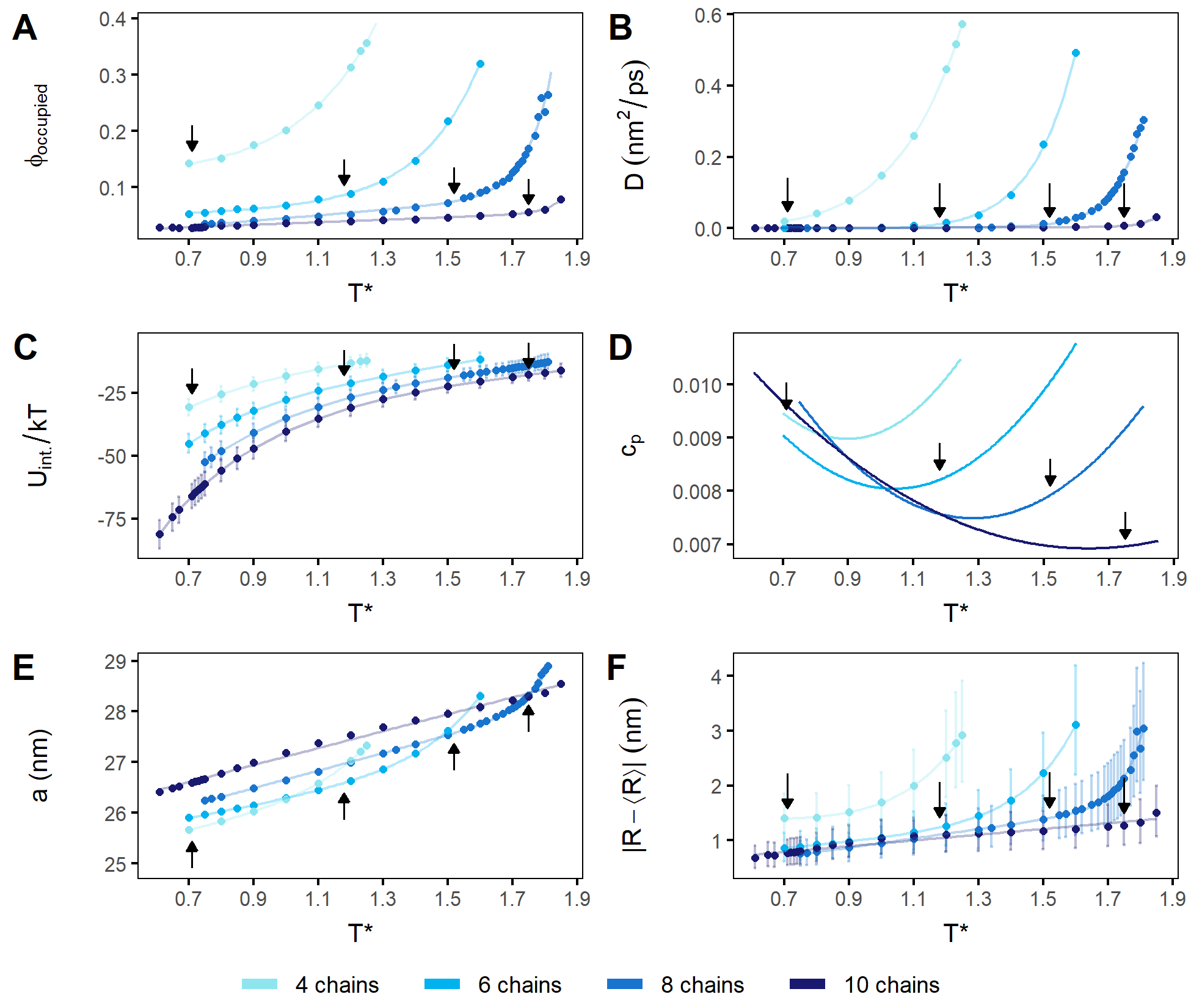}
    \caption{\label{fig:delocalization_all} Lattice properties as a function of temperature, for all systems studied. Fits are included as visual guides. Black arrows indicate $T_\text{deloc}$, the onset of delocalization for each system; see SI for how these are identified. Note that all properties change linearly with temperature below and exponentially above $T_\text{deloc}$, with the exception of $U_\text{int}$.
    \textbf{(A)} Approximate volume fraction occupied by 70\% of the small particles' probability $\phi_\text{occupied}$. This is a qualitative measure of delocalization.
    \textbf{(B)} Diffusion constant $D$ of the small particles.
    \textbf{(C)} Average interaction energy between small and large particles $U_\text{int}$, per small particle.
    \textbf{(D)} Specific heat at constant pressure $c_p = C_p/N$ ($N$ is the total number of particles) of the system. These curves were calculated by fitting spline curves to mean values of enthalpy and differentiating those curves with temperature.
    \textbf{(E)} Average BCC lattice parameter $a$.
    \textbf{(F)} Lattice fluctuations, measured by the median displacement of large particles from their average positions. Uncertainty bars here indicate first and third quartiles, rather than standard deviation, due to the skewed nature of the underlying distribution.}
\end{figure}

Figure \ref{fig:loc-vs-deloc_deloc} shows the average visitation frequency of the small particle centers in one BCC unit cell, when localized and when delocalized. Though the system pictured has 6 chains per small particle, we see similar behavior for all values of chains per small particle: when localized, small particles occupy the tetrahedral sites, and when delocalized, they occupy a much larger volume. Even when delocalized, the small particles concentrate around the tetrahedral sites and form a pattern in which the additional volume occupied by the small particles is roughly along the edges of the BCC's Wigner-Seitz cell. This permits the small particles to move between nearest tetrahedral sites along an energetically-favorable path, equidistant to multiple neighboring lattice points. 

We observe a smooth change from localized-to-delocalized behavior in all cases. Both the onset of delocalization $T_\text{deloc}$ and the overall melting temperature of the lattice $T_\text{melt}$ are higher with more grafted chains per small particle; see Table S1. In these systems, the total strength of interaction between the small and large particles scales with the number of grafted chains. Therefore, we use the number of grafted chains per small particle and interaction strength interchangeably throughout this paper. Additionally, $T_\text{deloc}$ approaches $T_\text{melt}$ with increasing interaction strength, which means that we observe a range of behavior. In systems with 4 chains per small particle, $T_\text{deloc}$ is very low, and the sublattice is delocalized at almost every reported temperature. For 6 and 8 chains per small particle, $T_\text{deloc}$ is higher and the system is localized at low temperatures and delocalized at high temperatures. For 10 chains per small particle, $T_\text{deloc}$ is almost equal to $T_\text{melt}$, and the small particles exhibit almost no sublattice delocalization until just before crystal melting. 

In Figure \ref{fig:delocalization_all}, we plot structural and thermodynamic properties of the crystals to characterize their transition. These properties and their importance are listed below.

\begin{itemize}
    \item Occupied volume fraction of the small particles, $\phi_\text{occupied}$ (Figure \ref{fig:delocalization_all}A), is a qualitative metric that directly measures delocalization. It represents the approximate volume occupied by 70\% of the small particles' probability, as a fraction of the total available volume (see Supplementary Information for more information). In a previous study,\cite{Girard2019} delocalization was quantified using metallicity, a parameter associated to the Shannon entropy of the sublattice. Here we use a more direct parameter to measure the filling of space by small particles in the sublattice.
    \item Diffusion coefficient of the small particles, $D$ (Figure \ref{fig:delocalization_all}B), has been used to categorize the order of superionic transitions.\cite{Hull2004} We have observed that the localized-to-delocalized change is associated with both static properties like $\phi_\text{occupied}$ and dynamic properties like $D$.
    \item Average interaction energy of a small particle, $U_\text{int}$ (Figure \ref{fig:delocalization_all}C), is capable of reflecting structural changes.
    \item Specific heat capacity, $c_p$ (Figure \ref{fig:delocalization_all}D), provides insight into the order of phase transitions.
    \item Lattice parameter, $a$ (Figure \ref{fig:delocalization_all}E), had been shown to reflect a first order phase transition in previous work in charged colloidal systems.\cite{Lin2020}
    \item Median lattice fluctuations (Figure \ref{fig:delocalization_all}F) are essential for quantifying melting  through the Lindemann criterion.   
\end{itemize}
All properties are plotted as a function of reduced temperature $T^*$, which is the value of $k_BT$ in energy units. In each panel in Figure \ref{fig:delocalization_all}, a black arrow indicates the approximate $T_\text{deloc}$ for each system. This temperature is estimated from the diffusion properties of the sublattice; see SI for more information about how this was calculated.

Many of the properties in Figure \ref{fig:delocalization_all} exhibit two trends, one during and another before delocalization. The occupied volume fraction $\phi_\text{occupied}$, diffusion coefficient of the small particles $D$, the lattice parameter $a$, and the lattice fluctuations (Figures \ref{fig:delocalization_all}A, \ref{fig:delocalization_all}B, \ref{fig:delocalization_all}E, and \ref{fig:delocalization_all}F) all increase linearly below $T_\text{deloc}$ and exponentially above, until the lattice melts. These phenomena appear correlated; particles begin to both diffuse and occupy a larger volume at the same temperatures, which is also the point at which lattice expansion and lattice fluctuations begin to increase dramatically. These ties will be explored in later sections. 

The smooth increase in $\phi_\text{occupied}$, $a$, and other properties suggests that the change from localized-to-delocalized small particles is not a phase transition. This is corroborated by the behavior of the specific heat of the system, $c_p$, shown in Figure \ref{fig:delocalization_all}D. We observe that $c_p$ of all systems is continuous and convex, indicating that no phase transition occurs during the process of delocalization. This is expected because the change from localization-to-delocalization does not reflect a change in the BCC symmetry imposed by the large particles and so the small particles' energy landscape is not qualitatively impacted. 

Even though the $c_p$ curves do not exhibit evidence of a phase transition, they provide information about the underlying energy landscape of the system. We explain the convexity of the $c_p$ curves with the deactivation and activation of degrees of freedom into which energy can be distributed. The low temperature negative slope of these curves relates to the flattening of the local minima of the energy landscape. This flattening decreases the interactive ends' available configurational phase space, decreasing $c_p$. This is also why the slope is more negative for systems with higher interaction strength. At higher temperatures, new energy modes are enabled in the form of diffusion of the small particles and lattice vibrations. This eventually leads to delocalization, and $c_p$ continues to increase until the lattice fully melts. 

Notably, while the $\phi_\text{occupied}$ and $a$ change rapidly above $T_\text{deloc}$, the interaction energy does not. Figure \ref{fig:delocalization_all}C shows the average "binding energy" (the energy of interaction between the large particles and interactive ends, relative to when they are infinitely far apart) per small particle in the system as a function of temperature. That this quantity increases only linearly even above $T_\text{deloc}$ indicates that entropy plays an important role in delocalization. This will be discussed in Section \textit{\nameref{sect:theory}}.

\subsection{Lattice fluctuations are essential for delocalization}
\label{sect:fixed_lattice}

\begin{figure}
    \centering
    \includegraphics{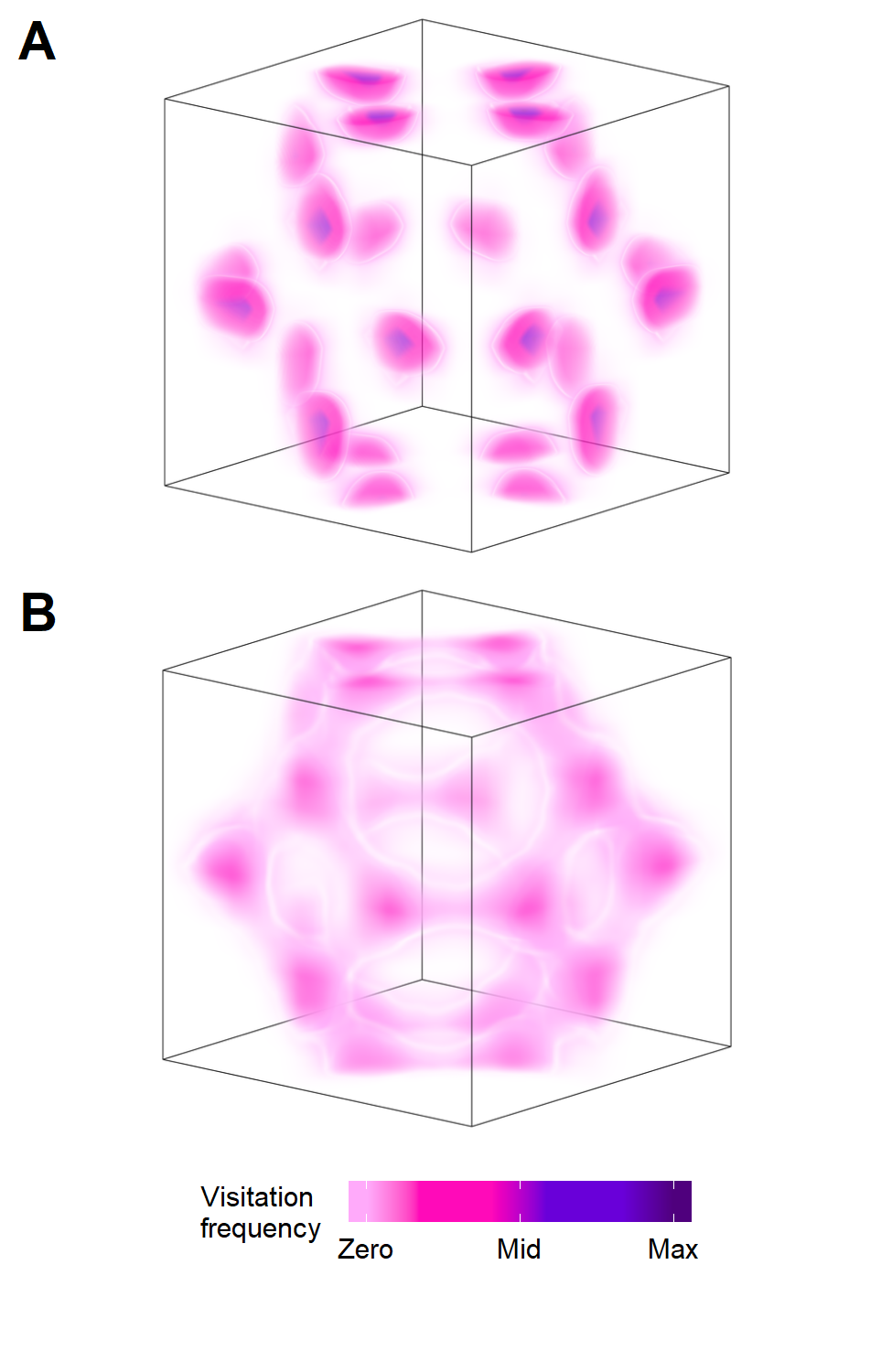}
    \caption{\label{fig:fixedlatt_loc-vs-deloc-images} The average visitation frequency of the small particle centers in a fixed lattice system with 6 chains per small particle. The maximum of the visitation frequency is 0.0250. Without lattice fluctuations, the small particles in the delocalized case occupy less volume than when lattice fluctuations are present. Note that the unit cells of these lattices are actually different sizes, but the images have been scaled such that the two are comparable.
    \textbf{(A)} $T^* = 0.9$. Small particles are localized on the tetrahedral sites of the BCC lattice. This is similar to the unconstrained lattice case.
    \textbf{(B)} $T^* = 1.6$. Small particles are delocalized. Again, they favor tetrahedral sites but also diffuse between sites.
    }
\end{figure}

To determine the importance of lattice fluctuations to delocalization, we ran additional simulations in which the large particles were fixed on their lattice points and not allowed to vibrate. The lattice parameter used for a given "fixed lattice" run was the mean value calculated from the unconstrained simulation with the same temperature and number of chains per small particle (Figure \ref{fig:delocalization_all}E). We found that without lattice vibrations, the small particles are not able to fully delocalize. This can be seen in the average visitation frequency plots in Figure \ref{fig:fixedlatt_loc-vs-deloc-images}. This is quantified by a large reduction in occupied volume fraction and a slight decrease of the diffusion coefficients relative to the unconstrained cases. This is similar to the finding by Schommers,\cite{Schommers1977} who saw diffusion in molecular dynamics models of superionic $\alpha$-AgI only when the iodine ion lattice was allowed to vibrate.  

\begin{figure}
    \centering
    \includegraphics{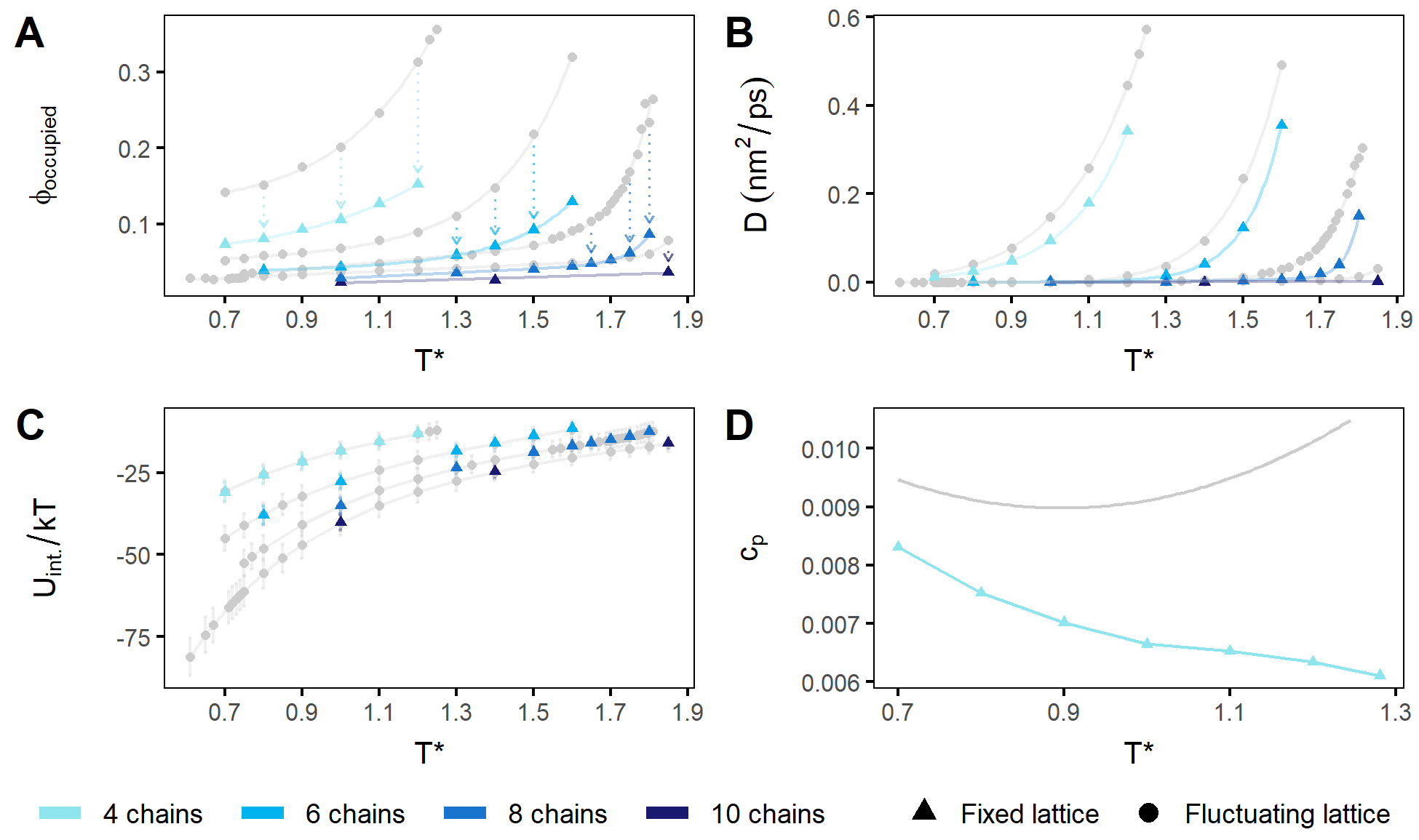}
    \caption{\label{fig:fixed_delocalization_all} Lattice properties as a function of temperature, for fixed lattice runs (compared to unconstrained runs). Data from the main runs (also in Figure \ref{fig:delocalization_all}) is shown in grey circles, and data from the fixed lattice runs is shown in blue triangles. Removing lattice fluctuations substantially suppresses delocalization and minorly suppresses diffusion.
    \textbf{(A)} Approximate occupied volume fraction $\phi_\text{occupied}$. Arrows connect fixed and fluctuating lattice simulations with the same number of chains per small particle as a visual guide. 
    \textbf{(B)} Diffusion constant $D$ of small particles. 
    \textbf{(C)} Average interaction energy $U_\text{int}$ of small particles with large particles. Because corresponding fixed and fluctuating lattice runs have the same average lattice constant, the average interaction energy of the small particles does not change, though the fluctuations of $U_\text{int}$ do.
    \textbf{(D)} The specific heat at constant pressure $c_p$ of the system with 4 chains per small particle.
    }
\end{figure}

Based on these results, delocalization is driven by both lattice vibrations and diffusion. We posit that vibration-driven delocalization occurs when lattice deformation either shifts the energy landscape sufficiently such that small particles can more easily diffuse, or that large particles pull small particles between tetrahedral sites while vibrating. Vibration-driven delocalization is fully suppressed in the fixed lattice simulations; this can be seen in Figure \ref{fig:fixed_delocalization_all}A. However, some delocalization remains due to small particle diffusion. As can be seen in Figure \ref{fig:fixed_delocalization_all}B, diffusion is still present in the fixed lattice simulations and appears to primarily depend on temperature and the lattice parameter, because they determine the flatness of the energy landscape. 

Analysis of the fixed lattice simulations demonstrates that delocalization is fully achieved only when both lattice vibrations and diffusion are present. The similarity between $c_p$ curves for the fixed and fluctuating lattice runs, shown in Figure \ref{fig:fixed_delocalization_all}D for the 4 chain system, underscores the importance of diffusion. With or without lattice vibrations, $c_p$ is continuous. Both $c_p$ curves exhibit an initial decrease characteristic of the flattening of the energy landscape but differ at higher temperatures. This is due to the lack of lattice vibrations in the fixed lattice simulations. As stated in the previous section, energy modes associated to the lattice vibrations are what drive the increase of $c_p$ after the flattening of the energy landscape. Therefore, $c_p$ for the fixed lattice simulations continues to decrease, whereas, the unconstrained simulations' $c_p$ increases.

\subsection{Vibrational entropy drives lattice expansion}
\label{sect:theory}

\begin{figure}
    \centering
    \subfloat{
        \includegraphics{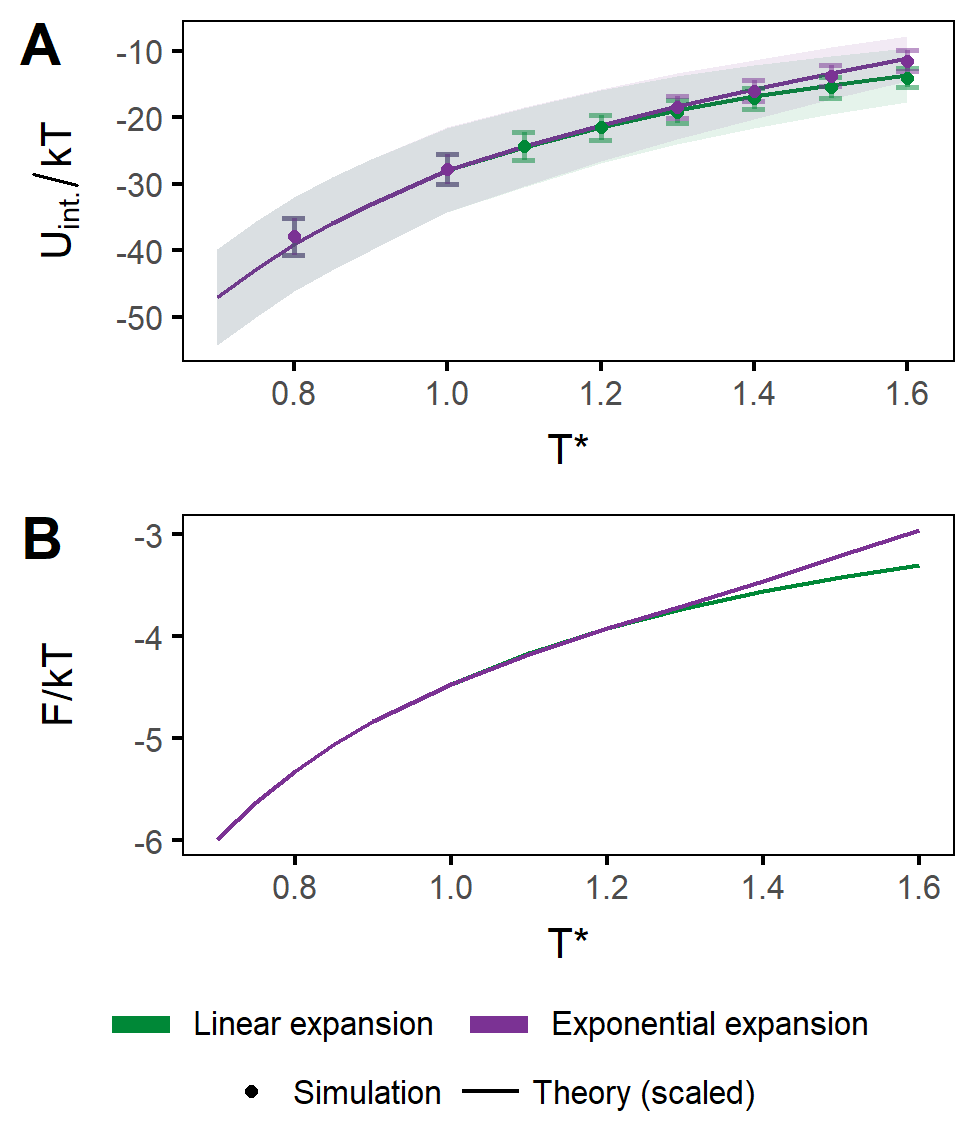}
    }
    \caption{\label{fig:theory_all} Comparison between theory and simulation (6 chain system).
    \textbf{(A)} Interaction energy between small and large particles. Line represents scaled theory results; points represent simulation results.
    \textbf{(B)} Theoretical prediction of the free energy of the systems with exponentially expanding lattice (main cases) and linearly expanding lattice. The free energy of the linear case is lower, indicating that something outside the theory must explain why the lattice expands exponentially.
    }
\end{figure}

Having established that lattice vibrations are crucial for delocalization, we turn to address the exponential expansion shown in Figure \ref{fig:delocalization_all}E. The exponential lattice expansion appears to be highly correlated with delocalization, but the reason that it occurs is unclear. To gain a better understanding, we performed a free energy analysis of our system using the same simplified theoretical model that predicted the energy landscape of a BCC unit cell in Figure \ref{fig:BCC_sites}B, and which is described in Section \textit{\nameref{sect:methods}}. This theoretical model describes one interactive end in a fixed (non-fluctuating) BCC unit cell of large particles. The energy of one lattice configuration based on the temperature, lattice parameter, and position of the interactive end $U_\text{end}(\vec{r},\vec{p};a)$ is found in Equation \ref{eq:theory-int-energy}, and is based on Equation \ref{eq:pair-potential} and Figure \ref{fig:model}A. Using these definitions and $a$ and $T$ from simulation, we calculated the partition function $Z(a, T) = \int e^{- U_\text{end}(\vec{r},\vec{p};a)/k_BT}~d\vec{r}~d\vec{p}~$ by numerically integrating over a unit cell. From this, we could calculate all relevant thermodynamic properties. See Section \textit{\nameref{sect:methods}} for more information.

We employed this model to explain why the lattice expands so rapidly at the onset of delocalization. To do this, we compared two cases: \latin{(i)} exponential expansion, which is the observed behavior of the lattice, and \latin{(ii)} linear expansion, in which the lattice expands only linearly over the entire temperature range. We ran fixed lattice simulations of both cases and compared those to theory.

Figure \ref{fig:theory_all}A shows the average interaction energy per small particle in the exponentially and linearly expanding cases, for simulation (points) and theory (solid line). We find that the theoretical model accurately predicts the energy in both cases up to a multiplicative factor. In the theoretical model, the interactive energy of a single interactive end is $\left< U_\text{potential} \right> = \frac{1}{Z(a, T)} \int U_\text{potential}(\vec{r}; a)~e^{- U_\text{end}(\vec{r}, \vec{p};a)/k_BT}~d\vec{r} d\vec{p}$ (a summary of Equation \ref{eq:theory-int-energy}). Because the result of this integral is the potential energy of one interactive end, we multiply $\left< U_\text{potential} \right>$ by the number of chains per small particle to estimate the total energy of a small particle. For example, Figure \ref{fig:theory_all} shows simulation results for runs with 6 chains per small particle. If the theoretical model were exact, we would multiply $\left< U_\text{potential} \right>$ by 6. However, the theoretical model overestimates the average energy per small particle relative to simulation. This is due to the fact that, in simulation, the limited reach of the chains does not allow the interactive end to fully explore the lowest energy portions of the cell's potential wells (this can be seen in the comparison between Figures \ref{fig:BCC_sites}B and  \ref{fig:BCC_sites}C). The result is that the interactive end's energy is about 20\% lower in simulation than in theory. Therefore, the theoretical results shown in Figure \ref{fig:theory_all}A are scaled by a factor of 0.82 (this factor differs by a few percent for the case of 4 chains per small particle). Additionally, excluded volume interactions of more densely grafted chains can impact the possible configurations of the interactive ends. This effect is not observed for small particles with 4 and 6 chains, because the average energy per chain is independent of the number of chains. Meanwhile, excluded volume interactions affect particle energy to a minor extent in systems with 8 and 10 chains per small particle.

The close correspondence between small particle potential energy in simulation and that predicted by theory indicates that the theoretical model can predict differences in properties between the exponentially and linearly expanding cases. Therefore, we used this model to compare the free energies of the two cases, to understand why one is favorable.
Using the partition function, we calculated the Helmholtz free energy, $F = -k_B T \ln Z$, which is plotted in Figure \ref{fig:theory_all}B. According to the theoretical model, the free energy of the linearly expanding lattice should be \textit{lower} than the free energy of the exponentially expanding lattice, so exponential expansion should be not favorable. 
We conclude, therefore, that at least one of the few interactions missing from the theoretical model must be what drives the observed exponential expansion. There are three pieces missing from the theoretical model: \latin{(i)} excluded volume interactions due to the presence of the other particle types, \latin{(ii)} bond constraints, and \latin{(iii)} lattice vibrations. We have already established that excluded volume interactions do not greatly impact the average energy of the small particles. Therefore, excluded volume should not contribute to the difference between the exponentially and linearly expanding cases, especially with 4 and 6 chains per small particle. We also can account for the bond constraints by scaling the potential energy by about 0.8, as mentioned above. Additionally, bond constraints limit the reach of the interactive ends and therefore are likely to make rapid lattice expansion less energetically favorable. Therefore, it must be vibrational entropy that drives the exponential lattice expansion. Additionally, large lattice fluctuations have already been seen to stabilize BCC crystals around their melting temperatures,\cite{Sprakel2017} which is possible due to BCC crystals' non-close packed structure and low coordination number. Vibrational entropy becomes dominant only above a certain temperature that depends on the number of chains per small particle. This is why we see exponential expansion and delocalization at different temperatures depending on the interaction strength. Based on this analysis, we can see that lattice vibrations determine both the degree of delocalization and the thermal expansion of the lattice.

\section{Conclusion}

In summary, we have seen that the localized-to-delocalized transition in 6:1 (BCC) binary colloidal systems is continuous, dominated by lattice vibrations, and tunable by number of chains per small particle. Our results suggest that the delocalization of the sublattice in this system is not a phase transition. This is supported by the fact that the symmetry of neither particle type changes during the transition; at all temperatures, the large particles form a BCC lattice and the small particles favor the BCC tetrahedral sites, even when delocalized.
The lack of a phase transition is also evidenced by the fact that $c_p$ is continuous for all systems. Moreover, delocalization is highly tied to vibrational entropy. Using simulations in which lattice vibrations were prohibited, as well as a free energy analysis with a simplified theoretical model, we conclude that most delocalization is driven by lattice vibrations, and that vibrational entropy is what causes the lattice to expand so rapidly above $T_\text{deloc}$. We can also see that the temperature range associated with the localized-to-delocalized transition is dependent on the number of chains per small particle, a proxy for interaction strength. The nature of the transition does not change as the number of chains per small particle does, but the presence of many chains suppresses delocalization almost entirely.  Additionally, the validation between the theoretical model and simulation results reveals that the potential energy landscape of a single interactive end within a BCC unit cell is a faithful representation of the simulated system, even though we have not included other particles within the theoretical model. This is accurate because of the asymmetry of range of interactions imposed by the size asymmetry of the particles.

Based on our analysis, we can identify additional conditions that appear to be favorable for sublattice delocalization. Our findings show that delocalization tends to occur at temperatures such that the small-large particle binding energy per chain is $\sim3-5~k_B T$. Per Figure \ref{fig:delocalization_all}C, delocalization occurs when the total interaction energy is around 22 $k_B T/$particle, distributed between all chains. This can be tuned by the number of chains grafted to the small particles. Additionally, we posit that, for the possibility of delocalization, the small particle size must be comparable to the fluctuations of the lattice and an asymmetry of interaction ranges must exist. The attraction range between small and large particles must be greater than the repulsion range between small particles because small particles must sit at and travel between interstitial sites that are closer together than lattice points. 
This asymmetry is present in our model, and it can also be achieved with charged colloids given a disparity in charge magnitudes between the small and large particles.\cite{Lin2020,Higler2017}. Note that increasing the range of repulsion between small particles may change the nature of the transition by adding correlations between small particles; however, we have not tested that here.

Under those conditions, similar analysis and conclusions may be generalized to other colloidal systems, with or without chains, but with certain caveats. For example, a more complex energy landscape with local energy minima at different interstitial symmetry points could change the nature of the transition. This could enable the small particles to transition through different symmetry points at different temperatures,\cite{Wang2015} which would be reflected in the order of the transition. This may be why previous work on charged colloidal systems\cite{Lin2020} found a discontinuity in certain physical parameters like the lattice constant. Additionally, other compositions produce different crystal types, and a phase transition between two crystal lattices can occur as a function of temperature. This can also impact the order of the localized-to-delocalized transition. Finally, the scaling of interaction strength with number of chains per small particle may not hold at system compositions that form non-BCC crystals, because the symmetry of collective chain configurations (impacted by the number of chains present) can affect the favorability of different interstitial points and crystal structures. 

Concerning the comparison between superionics and delocalized colloidal crystals, a continuous transition has also been reported between ionic and superionic states for some superionic crystals.\cite{Salamon1979} Therefore, drawing on the superionics literature can help us understand colloidal crystal delocalization and vice versa. For example, soft vibrational modes, which are high amplitude vibrations, are reported to be important for the presence of superionic conduction, most commonly a mobile cationic interstitial within an anionic lattice. Soft vibrational modes are stabilized by non-close packed crystals already seen in superionics\cite{Wang2015,Muy2018} as well as in these BCC colloidal crystals. The mechanism of this phenomenon is still not fully understood in superionic materials. However, it may be possible to use results reported here by drawing an analogy between the electron density's role in the stability of the crystal and that of the potential energy landscape of our system. The two may be compared by assuming polar covalent bonding between the static and mobile species. If true, then our findings using these colloidal systems would translate to superionic materials which are relevant to applications for the improved design of solid-state batteries for energy storage.\cite{Goodenough2013,Bachman2016,Famprikis2019}

% supporting information

\section{Supporting Information}
Supporting information contains: the pair distribution functions over the studied temperature range; details of the determination of $T_{\text{deloc}}$; the calculation method for occupied volume and the heat capacity; and an analysis of nearest neighbor interactions (PDF). Videos of rotating unit cells of a localized and delocalized sublattice for both unconstrained and fixed lattice simulations are also available (.mp4 videos). 

\begin{acknowledgement}

This work was supported by the Center for Bio-Inspired Energy Science, an Energy Frontier Research Center funded by the US Department of Energy, Office of Science, Basic Energy Sciences under Award DE-SC0000989. H.L.-R. thanks a fellowship from Fulbright-Garcia Robles and A.E. thanks a fellowship from the National Science Foundation under grant DGE-1450006. M.O.d.l.C. thanks the computational support of the Sherman Fairchild Foundation.

\end{acknowledgement}

\bibliography{refs}

\end{document}